# Leveraging Text and Knowledge Bases for Triple Scoring: An Ensemble Approach

## The BOKCHOY Triple Scorer at WSDM Cup 2017

Boyang Ding, Quan Wang*, Bin Wang
Institute of Information Engineering, Chinese Academy of Sciences, Beijing 100093, China
University of Chinese Academy of Sciences, Beijing 100049, China
{dingboyang,wangquan,wangbin}@iie.ac.cn

## ABSTRACT

We present our winning solution for the WSDM Cup 2017 triple scoring task. We devise an ensemble of four base scorers, so as to leverage the power of both text and knowledge bases for that task. Then we further refine the outputs of the ensemble by trigger word detection, achieving even better predictive accuracy. The code is available at https://github.com/wsdm-cup-2017/bokchoy.

## 1. INTRODUCTION

The WSDM Cup 2017 triple scoring task[1] is to compute relevance scores for triples from type-like relations [8]. Given such a triple, the score measures the degree to which the entity belongs to a specific type. For instance, the triple (JohnnyDepp, profession, Actor) may get a high score since acting is Depp's main profession, but (QuentinTarantino, profession, Actor) a low score since Tarantino is more of a director than an actor. Such scores are extremely useful in entity search [1].

The task was first recognized in [1], where a variety of methods have been proposed and tested. All these methods share a similar idea, i.e., to find "witnesses" for each triple in a text corpus, more specifically, Wikipedia. Take (JohnnyDepp, profession, Actor) for example. To score this triple, previous methods identify, from Depp's Wikipedia page, occurrences of words that are semantically related to the profession, i.e., witnesses. The more witnesses there are, the higher the score will be. Although such methods are generally reasonable and achieve relatively good performance, they still have limitations.

• Witnesses are collected only in Wikipedia, but not in the knowledge base where a triple comes from. This knowledge base itself, however, might contain rich evidence for that triple. For example, if we observe (JohnnyDepp, bornIn, Kentucky) and (Kentucky, locatedIn, US) in the knowledge base, we might probably assign a high score to the triple (JohnnyDepp, nationality, US).

• To collect witnesses, sentences of a Wikipedia page are treated equally, ignoring the order in which they appear. However, the first sentence is usually most informative for type-like relations. Take Depp's Wikipedia page for example. The first sentence

⌈*John Christopher "Johnny" Depp II (born June 9, 1963) is an American actor, producer, and musician.*⌋

indicates his nationality US and main profession Actor. Witnesses found in such sentences will definitely provide more confidence in judging triple scores.

---
*Corresponding author: Quan Wang (wangquan@iie.ac.cn).
[1]http://www.wsdm-cup-2017.org/triple-scoring.html

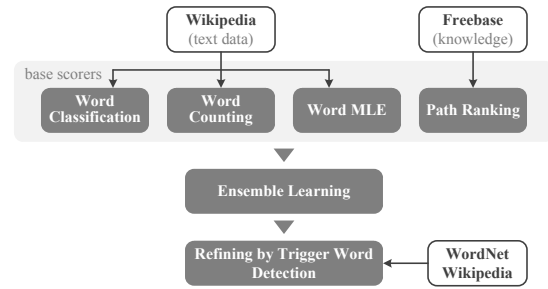

**Figure 1: An overview of the BOKCHOY triple scorer.**

• Witnesses are found by detecting words that are semantically related to a specific type. Measuring semantic relatedness, however, has long been considered a challenging task in natural language processing. Even with the help of topic modeling techniques [4, 9] (as investigated in [1]), it might still be hard to measure semantic relatedness accurately. Consider, for example, the profession Athlete. It is almost impossible to make every judgement correct whenever a word such as ⌈*runner*⌋, ⌈*jumper*⌋, or ⌈*swimmer*⌋ is detected.

To overcome these limitations, we devise an ensemble model for triple scoring, named BOKCHOY. The overall framework of our approach is sketched in Figure 1. As base scorers, we employ word classification [1], word counting [1], word MLE [1], and also path ranking [10]. The first three base scorers find witnesses on the basis of Wikipedia, while the last one further makes use of Freebase [5]. The base scorers are then combined into an ensemble by weighted averaging [15]. After that, an elongation step is introduced to further refine outputs of the ensemble scorer. Specifically, we create, for each target type, a list of trigger words by using publicly available lexical resources like WordNet [11]. Trigger words of the profession Athlete, for example, may include hyponyms of ⌈*athlete*⌋, such as ⌈*runner*⌋ and ⌈*jumper*⌋. Given a triple, we detect, from the first sentence of the entity's Wikipedia page, occurrences of such trigger words, and accordingly refine the output of the ensemble.

Our main contributions can be summarized as follows.

• Besides the methods introduced in [1], we further employ path ranking as a base scorer. As such, we can find witnesses not only in text data, but also in knowledge bases.

• We use ensemble learning to combine multiple base scorers, achieving better predictive performance than each single model.

• We propose to further refine the ensemble scorer by trigger word detection. Trigger words are extracted from lexical resources such as WordNet, and detection is confined only to the first sentence of a Wikipedia page. Both are more indicative of a given type.

## 2. TASK DESCRIPTION

The WSDM Cup 2017 triple scoring task is to predict, for each triple from type-like relations, a relevance score in the range of 0-7. Two types of such relations are considered, i.e., `profession` and `nationality`. The prediction task is confined to 385,426 different persons, 200 different professions, and 100 distinct nationalities, all contained in Freebase. Let $\mathcal{E}$, $\mathcal{P}$, and $\mathcal{N}$ denote the sets of persons, professions, and nationalities, respectively. Training data includes:

- `wiki-sentences`: 33,159,353 sentences from the English version of Wikipedia with annotations of the 385,426 persons in $\mathcal{E}$;
- `profession.kb`: all professions for a subset of 343,329 persons, extracted from a 14-04-2014 dump of Freebase;
- `nationality.kb`: all nationalities for a subset of 301,590 persons, extracted from the same dump;
- `profession.train`: relevance scores for 515 triples (pertaining to 134 persons) from `profession.kb`;
- `nationality.train`: relevance scores for 162 triples (pertaining to 77 persons) from `nationality.kb`.

Submissions are evaluated on a test set consisting of triples from the two *.kb files, but with relevance scores as ground truth (not released to participants). Three metrics are used for evaluation, i.e., accuracy (ACC), average score difference (ASD), and Kendall's tau (TAU). The award goes to the submission with the highest ACC. For more details about the task, please refer to [2].

## 3. OUR APPROACH

As illustrated in Figure 1, our approach is an ensemble of four base scorers, where the first three leverage the power of text in Wikipedia, and the last the power of knowledge in Freebase. Outputs of the ensemble are further refined by trigger word detection, so as to get even better predictive accuracy. Our approach does not require manually labeled data (i.e., relevance scores for triples). The two *.train files are used only as development sets to guide the design of our approach. In what follows, we describe the key components of our approach, including base scorer learning, ensemble learning, and refining by trigger word detection.

### 3.1 Base Scorer Learning

This section describes our four base scorers. For illustration convenience we describe these models using the `profession` relation. But they also work with the `nationality` relation as well.

#### 3.1.1 Leveraging Text in Wikipedia

Wikipedia text contains rich information about persons' professions and nationalities. A variety of methods based on Wikipedia have been proposed and found useful in triple scoring. We use three such methods, i.e., word classification [1], word counting [1], and word MLE [1] as our base scorers.

**Text Training Data** We follow [1] and adopt a similar heuristic to obtain labeled training examples for these base scorers. For each profession in $\mathcal{P}$, we select, from the `profession.kb` file, people with only that profession as *positive candidates*, and people who do not have that profession at all as *negative candidates*. Then we associate, with each person in $\mathcal{E}$, all sentences in the `wiki-sentences` file that get at least one mention linked to that person. We call these sentences the person's *associated text*. Stop words in a standard list provided in `scikit-learn` [12] are further removed.

**Word Classification** The first base scorer trains for each profession a binary classifier to judge whether that profession is primary or not for a person, according to his/her associated text. For each profession in $\mathcal{P}$, we train the classifier with labeled examples sampled from the positive and negative candidates. We employ a similar sampling strategy as practiced in [1] to get a representative set of persons with all levels of *popularity*. Here, popularity is the number of times a person is mentioned in `wiki-sentences`. Specifically, we define buckets of popularity $[2^i, 2^{i+1})$ for $0 \le i \le \lfloor \log_2 P \rfloor$, where $P$ is the maximal popularity and $\lfloor \cdot \rfloor$ means round down to the nearest integer. Then we sample uniformly from each bucket at most 100 positive candidates (i.e., *positive examples*), and the same number of negative candidates (i.e., *negative examples*).[2] These examples are used to train the classifier. We hope this sampling strategy can make the distribution of persons in training data similar to that in test data.

Given a positive or negative example, we use words in the associated text as features. Feature values are calculated by their tf-idf values [13] in the training corpus. Here, the training corpus consists of text associated with all the *positive and negative examples*. To speed up training, we perform feature selection. Only the top 20,000 words with highest frequencies in the training corpus are selected. We use the `LogisticRegressionCV` tool[3] provided in `scikit-learn` to train an $\ell_2$-regularized logistic regression classifier. We choose the `liblinear` solver, and conduct 5-fold cross-validation to select the optimal regularization parameter in a logarithmic scale from $10^{-4}$ to $10^4$. Other parameters are set to their default values. During prediction, we construct for each person a feature vector using his/her associated text, and define the relevance score as the confidence value predicted by the learned classifier.

**Word Counting** The second base scorer takes words as indicators of professions, and predicts a person's main profession by judging how much his/her associated text is indicative of that profession. To do so, for each profession in $\mathcal{P}$, we construct a training corpus consisting of text associated with only the *positive candidates*. We then compute the tf-idf value for each word in the training corpus, and weight that word by its tf-idf value. To speed up the learning process, we consider only the top 100,000 words with highest frequencies in the corpus. During prediction, we compute for a given person the relevance score as $s = \sum_{w_i} n_{w_i} \cdot tf\text{-}idf_{w_i}$, where $n_{w_i}$ is the number of times the word $w_i$ occurs in the person's associated text and $tf\text{-}idf_{w_i}$ is the weight of that word.

**Word MLE** Our third base scorer is a generative model where a person's associated text can be generated from his/her professions. Given a person with $k$ professions and $n$ words in his/her associated text: pick a profession $p_i$ from the $k$ professions with probability $P(p_i)$; generate a word $w_j$ from that profession with probability $P(w_j|p_i)$; repeat until all the words are generated. The profession probability $P(p_i)$ can then be used to score triples, i.e., measuring the relevance of profession $p_i$ to that person.

These profession probabilities can be obtained by maximum likelihood estimation (MLE). The log-likelihood of generating the $n$ words from the $k$ professions is:

$$\log \mathcal{L} = \sum_{j=1}^{n} \Big[ tf_j \cdot \log \Big( \sum_{i=1}^{k} P(p_i) P(w_j|p_i) \Big) \Big],$$

where $tf_j$ is the frequency of word $w_j$, computed as its tf-idf value in the text associated with *all* persons in $\mathcal{E}$. Only the top 20,000 words with highest frequencies are kept for efficiency reasons. To compute $P(w_j|p_i)$, we collect text associated with the *positive candidates* for profession $p_i$, and calculate the tf-idf value of word $w_j$. We further follow [1] and add a pseudo profession $p_0$ to each person. We use text associated with 10,000 persons randomly selected from $\mathcal{E}$ to derive $P(w_j|p_0)$.[4] During prediction, we estimate for each person $\vec{p} = [P(p_0), \cdots, P(p_k)]$ that maximizes $\log \mathcal{L}$ with the constraint $\sum_i P(p_i) = 1$. We use the EM algorithm [6]. The E step and M step are identical to those in pLSI [9], but with fixed $P(w_j|p_i)$ values. Please refer to [1, 9] for more details.

---

[2] If a bucket contains fewer than 100 positive candidates, we take all of them from that bucket. Since there are much more negative candidates than positive ones, we can always sample the same number of positive and negative candidates from a bucket.

[3] http://scikit-learn.org/stable/modules/generated/sklearn.linear_model.LogisticRegressionCV.html

### 3.1.2 Leveraging Knowledge in Freebase

The methods described above exploit only Wikipedia text for the triple scoring task. In this section, we further introduce a new base scorer, i.e., path ranking [10], to leverage knowledge in Freebase for that task. Path ranking is an approach to reasoning on knowledge bases. The key idea is to build for each relation a binary classifier, with paths that connect two entities as features, to predict whether the two entities should be linked by that relation or not [7]. For example, `bornIn` → `locatedIn` is a path linking `JohnnyDepp` to `US` (through an intermediate node `Kentucky`). This path can be used as a feature to predict presence/absence of the relation `nationality` between the two entities. We can then score triples with outputs of such classifiers.

**Freebase Training Data** To obtain labeled training data for path ranking, we use the 11-10-2016 dump of Freebase,[5] and remove (i) triples that do not contain any person in $\mathcal{E}$, and (ii) triples from general relations such as `/base/*` and `/common/*`. In this manner, we obtain a subset of Freebase consisting of 10,743,200 entities, 3,285 relations, and 26,468,661 triples, referred to as `Freebase-person`. Given the target relation `profession`, we filter out persons with their professions observed in `Freebase-person`, rank them by the number of associated triples, and select top 10,000 persons (as well as their professions) as *positive examples*. For each positive example, i.e., a person with one of his/her professions, we construct a *negative example* by randomly replacing that profession with another one in $\mathcal{P}$. We further ensure that negative examples are observed neither in `Freebase-person` nor in `profession.kb`.

**Path Ranking** A typical path ranking algorithm consists of three steps, i.e., feature extraction, feature computation, and classification [16]. To extract features, given a labeled example, we conduct depth-first search to enumerate all paths with a bounded length of $\ell \leq 3$ between the two entities. We use the code provided in [14],[6] but the difference is that we do not block a specific relation during extracting path features for that relation. After feature extraction, we simply compute the value of each feature as the number of times it appears in each labeled example. Then we adopt `RandomForest Classifier`[7] provided in `scikit-learn` to train a binary classifier. We randomly split the labeled examples into 70% training and 30% validation. The parameter "n_estimators" (the number of trees in the forest) is set to 300, "min_samples_split" (the minimum number of samples required to split an internal node) tuned in {2,5,10}, and "max_features" (the number of features to consider when looking for the best split) in {"sqrt", "log2"}. Other parameters are set to their default values. The optimal configuration is determined by maximizing AUC-ROC on the validation set. During prediction, for each triple, we extract path features between the two entities, and

[4]We observed no significant improvements after adding the pseudo profession, so we did not do this for the `nationality` relation.
[5]https://developers.google.com/freebase/
[6]https://github.com/nddsg/KGMiner
[7]http://scikit-learn.org/stable/modules/generated/sklearn.ensemble.RandomForestClassifier.html

|  | profession | nationality |
|---|---|---|
| Word Classification | Mapscale | Maplog |
| Word Counting | Maplin | Maplin |
| Word MLE | Maplog | Maplog |
| Path Ranking | Mapscale | Maplog |

**Table 1: Mapping strategies used in the four base scorers.**

score that triple with the class probability predicted by the learned classifier.[8]

### 3.1.3 Mapping to Triple Scores

The above base scorers yield a variety of results, e.g., confidence values, weighted sums, and probabilities. We employ three strategies to map such results to integer triple scores in the range of 0-7.

**Maplin** Maplin is a linear strategy proposed in [1], mapping a value $s$ to a triple score $s'$ as:

$$s' = \lfloor \tfrac{s}{s_{\max}} \times 7 \rfloor,$$

where $s_{\max}$ is the highest value that a person get for all his/her professions, and $\lfloor \cdot \rfloor$ means round down to the nearest integer.

**Maplog** Maplog is also proposed in [1]. It is a mapping in a logarithmic scale, defined as:

$$s' = \lfloor \max\{0, \log_2(\tfrac{s}{s_{\max}} \times 2^7)\} \rfloor.$$

Note that we might get $\log_2(\tfrac{s}{s_{\max}} \times 2^7) < 0$ for an $s$ small enough. We set $s' = 0$ for such $s$ values.

**Mapscale** Besides Maplin and Maplog, we design another strategy Mapscale. It is a linear mapping applied only on probabilities:

$$s' = \lfloor s \times 8 - \epsilon \rfloor,$$

where $\epsilon = 10^{-4}$ so that we can get $s' = 7$ with $s = 1$.

Table 1 summarizes the mapping strategies used in the four base scorers for each of the two relations. These are the optimal choices that yield the highest ACC values on the two *.train files.

## 3.2 Ensemble Learning

After we obtain the four base scorers, we combine them into an ensemble to achieve better predictive performance. Here we choose weighted averaging [15], which defines a triple $t$'s relevance score $S(t)$ as:

$$S(t) = \lfloor \sum_{i=1}^{M_t} w_i s_i(t) \rfloor.$$

Here, $M_t$ is the number of base scorers that are used to score the triple;[9] $s_i(t)$ an integer relevance score in the range of 0-7 predicted by the $i$-th base scorer; and $w_i = \text{ACC}_i / \sum_{j=1}^{M_t} \text{ACC}_j$ the weight of that base scorer. $\text{ACC}_j$ is the ACC value that the $j$-th base scorer yield on the corresponding *.train file.

[8]There are too many test triples and extracting path features for all of them can be extremely time-consuming. So, for the `profession` relation, we consider only persons with at least four different professions and extract path features accordingly. That means, triples associated with the other persons are scored only by the first three base scorers, without path ranking. We do not use such filtering for the `nationality` relation.
[9]Given a triple, there is a chance that some of the four base scorers (or even all of them) cannot output a relevance score (recall path ranking on the `profession` relation). We set $S(t) = 0$ if $M_t = 0$, i.e., all the base scorers fail to make a prediction.

## 3.3 Refining by Trigger Word Detection

Finally, considering that sentences at the very beginning of one's Wikipedia page are usually more indicative of his/her main profession and nationality, we further detect from such sentences *trigger words* of professions and nationalities, and refine outputs of the ensemble accordingly. Given a specific type, i.e., a profession or nationality, trigger words are those that have the same meaning with the type (or any specialization of it). Trigger words of a profession include (i) the original and plural forms, e.g., ⌈actor⌉ and ⌈actors⌉ for `Actor`, (ii) synonyms, e.g., ⌈enterpriser⌉ for `Entrepreneur`, and (iii) hyponyms, e.g., ⌈runner⌉ and ⌈jumper⌉ for `Athlete`. Synonyms and hyponyms are obtained from WordNet 3.0.[10] Trigger words of a nationality include (i) the country name and (ii) its adjectival form, e.g., ⌈Germany⌉ and ⌈German⌉ for `Germany`. Country names and adjectival forms are collected from a publicly available resource.[11] Besides these, we manually create 25 trigger words for some nationalities, e.g., ⌈British⌉ for `UnitedKingdom`. The whole list is available along with our source code.[12] After creating trigger words, we associate with each person a short description indicated by the Freebase relation `common/topic/description`. This actually is the first paragraph of the person's Wikipedia page. The first sentence of the description is further recognized with the tokenizer provided in Natural Language Tookit (NLTK) [3].[13]

Then, given a triple stating that a person belongs to a type, we detect occurrences of trigger words of that type (exact string match) from the person's description, and accordingly refine the relevance score of the triple output by the ensemble: (i) if at least one trigger word has been detected in the *first sentence* and the relevance score is lower than 5, upgrade it to 5; (ii) if none of the trigger words has been detected in the *description* and the relevance score is higher than 2, degrade it to 2.[14] That means, professions or nationalities mentioned in the first sentence of a person's Wikipedia page are usually taken as his/her main profession or nationality, while those not mentioned in the first paragraph are probably not.

## 4. EVALUATION RESULTS

In this section we present experiments and results on the two `*.train` files. For a fair comparison, we select from the two files triples that can be predicted by all the four base scorers, i.e., 485 out of 515 triples from `profession.train` and 160 out of 162 triples from `nationality.train`. Note that this is not the final test data. We just use it to verify the effectiveness of each component of our approach, i.e., base scorers, different ensemble strategies, and refinement carried out by trigger word detection.

**Base Scorers** We first test the performance of using each of the four base scorers alone. The results are shown in Table 2 (the first part). We can see that (i) Word Classification and Word Counting perform quite well on both relations, but Word MLE substantially worse than them; (ii) Path Ranking performs best on `nationality`, but not well enough on `profession`. The reason may be that paths predictive for different professions are much more diverse than those predictive for different nationalities. For example, `castIn` is indicative only of `Actor`, but not other professions such as `Engineer` and `Farmer`. Building a separate classifier for each profession may be a better choice than mixing them together.

---

[10] https://wordnet.princeton.edu/wordnet/download/
[11] http://siteresources.worldbank.org/ TRANSLATIONSERVICESEXT/Resources/ CountryNamesandAdjectives.doc
[12] Actually, there are other publicly available resources that can be used to define trigger words for nationalities, e.g., the one created by Wikipedia (available at https://en.wikipedia.org/wiki/List_of_ adjectival_and_demonymic_forms_for_countries_and_nations). This list is more comprehensive than the one we used in the task, and hence might not require manually creating trigger words.
[13] http://www.nltk.org/api/nltk.tokenize.html
[14] For the `profession` relation, we use only rule-i but not rule-ii. The reason is that a profession can have various specializations and generalizations, which is usually more difficult to detect.

|  | profession | | | nationality | | |
|---|---|---|---|---|---|---|
|  | ACC | ASD | TAU | ACC | ASD | TAU |
| Word Classification | 0.74 | 1.79 | 0.32 | 0.72 | 1.82 | 0.45 |
| Word Counting | 0.72 | 1.69 | 0.30 | 0.76 | 1.68 | 0.45 |
| Word MLE | 0.59 | 2.46 | 0.31 | 0.64 | 2.09 | 0.41 |
| Path Ranking | 0.67 | 2.06 | 0.40 | 0.76 | 1.57 | 0.44 |
| Ensemble | 0.75 | 1.63 | 0.27 | 0.78 | 1.54 | 0.39 |
| − Word Classification | 0.73 | 1.69 | 0.26 | 0.79 | 1.56 | 0.38 |
| − Word Counting | 0.74 | 1.66 | 0.27 | 0.78 | 1.56 | 0.37 |
| − Word MLE | 0.78 | 1.58 | 0.29 | 0.84 | 1.57 | 0.41 |
| − Path Ranking | 0.72 | 1.76 | 0.27 | 0.76 | 1.69 | 0.39 |
| **Ensemble (R)** | **0.84** | **1.44** | **0.25** | **0.89** | **1.27** | **0.30** |
| − Word Classification (R) | 0.83 | 1.47 | 0.25 | 0.90 | 1.25 | 0.28 |
| − Word Counting (R) | 0.83 | 1.44 | 0.25 | 0.89 | 1.27 | 0.29 |
| − Word MLE (R) | 0.81 | 1.52 | 0.28 | 0.91 | 1.34 | 0.29 |
| − Path Ranking (R) | 0.82 | 1.44 | 0.25 | 0.89 | 1.28 | 0.30 |
| TWD Alone | 0.51 | 2.65 | 0.45 | 0.86 | 1.76 | 0.27 |

Table 2: Evaluation results on the two `*.train` files.

**Ensemble Strategies** We further investigate different strategies to combine the base scorers into an ensemble. The results are given in the second part of Table 2, where "Ensemble" means combining all the four base scorers, and "Ensemble−Word MLE", for example, combining the other three base scorers except Word MLE. We can see that (i) Combining multiple models generally performs better than using a single model alone, and Ensemble gets relatively good performance among these strategies; (ii) Ensemble−Word MLE performs even better than Ensemble (in ACC), due to the low performance of Word MLE; (iii) Ensemble−Path Ranking performs almost the worst among these strategies. This is because Path Ranking, which leverages Freebase rather than Wikipedia text for triple scoring, is the most different base scorer from the others. Combining it into the ensemble can achieve maximum benefits.

**Refining by Trigger Word Detection** Finally, we test the effectiveness of refinement carried out by trigger word detection. We refer to a model with such refinement as, for example, "Ensemble (R)". Table 2 (the third part) shows the results. We can see that refining by trigger word detection always brings significantly better results, on both relations and with all the ensemble strategies.

We further test the performance of using trigger word detection alone, referred to as "TWD Alone". This approach scores a triple solely on the basis of trigger words, without using any base or ensemble scorers: (i) if at least one trigger word has been detected in the *first sentence*, give a score of 5; (ii) if none of the trigger words has been detected in the *description*, give a score of 2; and (iii) give a score of 3 or 4 with equal probability otherwise. The results are shown in Table 2 (the last part). We can see that using trigger word detection alone performs substantially worse than using it as refinement over ensemble scorers. This may be caused by the case where trigger words are detected in the *description* but not in the *first sentence*. In this case, TWD Alone has to make random guesses while ensemble scorers can still make predictions from other evidence. Furthermore, TWD Alone performs worse on `profession` than on `nationality`. This is because a profession can have various specializations and generalizations, which is usually more difficult to detect than a nationality.

We choose *Ensemble (R)* as our final solution, i.e., a combination of four base scorers refined by trigger word detection. The results on the test data are *0.87 in ACC*, *1.63 in ASD*, and *0.33 in TAU*.

## 5. CONCLUSION

We devise an ensemble of four base scorers for triple scoring, so as to leverage the power of both text and knowledge bases for that task. Compared with previous work, our solution is superior in that (i) we employ path ranking as a base scorer which can further leverage the power of Freebase; (ii) we use ensemble learning which can take advantage of various base scorers without overfitting; and (iii) we conduct trigger word detection in the very beginning of one's Wikipedia page and refine outputs of the ensemble accordingly.

## Acknowledgments

We thank the WSDM Cup 2017 organizers for a challenging and exciting competition. This work is supported by the National Natural Science Foundation of China (grant No. 61402465).